\documentclass[12pt,preprint]{aastex}
\usepackage{graphicx}
\usepackage{soul}
\usepackage{multirow}
\usepackage{makecell}
\usepackage{CJKulem}
\usepackage{color}
\shorttitle{}
\shortauthors{}

\begin{document}
\title{
Exploring the Cosmic Reionization Epoch in Frequency Space:
An Improved  Approach to Remove the Foreground in 21 cm Tomography
}
\author{Jingying Wang\altaffilmark{1}, Haiguang Xu\altaffilmark{1}, Tao An\altaffilmark{2},
  Junhua Gu\altaffilmark{3}, Xueying Guo\altaffilmark{1}, Weitian Li\altaffilmark{1},  Yu Wang\altaffilmark{2}, Chengze Liu\altaffilmark{1}, Olivier Martineau-Huynh\altaffilmark{3, 4}, and Xiang-Ping Wu\altaffilmark{3}}

 \altaffiltext{1}{Department of Physics, Shanghai
  Jiao Tong University, 800 Dongchuan Road, Minhang, Shanghai 200240,
  China; e-mail: hgxu@sjtu.edu.cn, zishi@sjtu.edu.cn;}

\altaffiltext{2}{Shanghai Astronomical Observatory, Chinese Academy of Sciences, Nandan Road 80, Shanghai, 200030, China;}
 \altaffiltext{3}{National Astronomical Observatories, Chinese Academy of Sciences, 20A Datun Road, Beijing 100012, China;}

\altaffiltext{4}{ Laboratoire de Physique Nucl\'{e}aire et de Physique des Hautes Energies,
CNRS/IN2P3 and Universit\'{e} Pierre et Marie Curie, Tour 12-22, 4 place
Jussieu, 75005 Paris, France.
}

\begin{abstract}
Aiming to correctly restore the redshifted 21 cm signals emitted by the neutral hydrogen during the cosmic reionization processes, we re-examine the separation approaches based on the quadratic polynomial fitting technique in frequency space to investigate whether they works satisfactorily with complex foreground, by quantitatively evaluate the quality of restored 21 cm signals in terms of sample statistics. We construct the foreground model to characterize both spatial and spectral substructures of the real sky, and use it to simulate the observed radio spectra. By comparing between different separation approaches through statistical analysis of restored 21 cm spectra and corresponding power spectra, as well as their constraints on the mean halo bias $b$ and average ionization fraction $x_e$ of the reionization processes, at $z=8$ and the noise level of 60 mK we find that, although the complex foreground can be well approximated with quadratic polynomial expansion, a significant part of Mpc-scale components of the 21 cm signals (75\% for $\gtrsim 6h^{-1}$ Mpc scales and 34\% for $\gtrsim 1h^{-1}$ Mpc scales) is lost because it  tends to be mis-identified as part of the foreground when single-narrow-segment separation approach is applied. The best restoration of the 21 cm signals and the tightest determination of $b$ and $x_e$ can be obtained with the three-narrow-segment fitting technique as proposed in this paper. Similar results can be obtained at other redshifts. 

\end{abstract}


\section{INTRODUCTION}

In the late stage of the dark ages of the Universe,
  neutral hydrogen in the intergalactic medium (IGM) began to be reionized by UV and soft X-ray photons (Baek et al. 2010) emitted by the first-generation   sources,  such as stars and/or quasars (see Morales \& Wyithe 2010 for a recent review). This period, which started sometime from about 150 Myrs to 500 Myrs since the Big Bang, is known as the epoch of reionization (EoR), and is one of  a few most important milestones in the evolution histories of both our Universe and galaxies.
The study of the line emission of neutral hydrogen at 21 cm (21 cm signals hereafter) coming from the surroundings of ionizing sources in EoR  provides us with a novel opportunity to reveal the properties of the first-generation stars and/or quasars, as well as to constrain cosmological models
(e.g., Iliev et al. 2002; Furlanetto et al. 2004a;  Santos \& Cooray 2006; Morales \& Wyithe 2010).
 EoR is expected to be a major center of interest within the next one or two decades for the facilities working in the low-frequency radio band ($\lesssim 1$ GHz), such as the 21 Centimeter Array (21CMA\footnote[1]{http://21cma.bao.ac.cn}), the Low Frequency Array (LOFAR\footnote[2]{http://www.lofar.org}),  the  Murchison Wide-field Array (MWA\footnote[3]{http://www.mwatelescope.org}), and the Square Kilometer Array (SKA\footnote[4]{http://www.skatelescope.org}).

One of the serious challenges in observing  the 21 cm signals from EoR is that they are extremely weak ($\sim 10$ mK). Having been redshifted to the low-frequency radio band ($50-200$ MHz, for $6<z<27$; Morales \& Wyithe 2010; Field 1959a, 1959b; Scott \& Rees 1990; Madau et al. 1997), the 21 cm signals are drowned in the luminous foreground, which consists of our Galaxy, extragalactic discrete sources (e.g., radio galaxies and active galactic nuclei, i.e., AGNs), galaxy clusters, and other relatively minor  contaminating  sources. Theoretical and  numerical efforts have been made in the  past few years to probe how to correctly restore the redshifted 21 cm signals from this overwhelmingly luminous  foreground,  e.g., extraction  of the 21 cm signals in either the angular power spectrum  space (e.g., Di Matteo et al. 2002, 2004; Santos et al. 2005; McQuinn et al. 2006), or the frequency space (e.g., Wang et al. 2006; Gleser et al. 2008; Jeli{\'c} et al. 2008; Bowman et al. 2009; Harker et al. 2009a; Liu et al. 2009; Petrovic \& Oh 2011), both on the {\it a priori} assumption that the
 integral foreground emission can be well described as a smooth continuum. 

For the separation works in the frequency space,  the total spectrum is usually fitted with a smooth function (e.g., an n-order polynomial model $\log {T}_{\rm fore}^{\,\prime}(\nu) = a_0+\sum_{i=1}^{N_{\rm poly}} a_i \log \nu^i$, where $i=1,~2,~...,~N_{\rm poly}$) to describe the integral foreground brightness temperature ${T}_{\rm fore}(\nu)$.
However, the first thing one shall keep in mind is that the foreground models adopted by previous works are relatively simple more or less, e.g., extragalactic discrete sources are treated as point sources, showing no substructure such as lobes and jets that are often observed in nearby FRI and FRII radio galaxies,
and/or possessing simple, featureless power-law radio spectra. Meanwhile, galaxy clusters, which contribute to the foreground at a non-negligible level compared with AGNs, are not taken into account, or treated as simple power-law spectrum sources.
 A revisit to the separation approach with more complex and more realistic foreground models is necessary.
Secondly, a variety of theoretical and  numerical works (e.g., Furlanetto et al. 2004b; Wyithe \& Loeb 2004; Furlanetto \& Oh 2005; Petrovic \& Oh 2011)  have already indicated that  in late phases of the cosmic reionization, the typical sizes of ionized regions are of the order of a few tens of comoving Mpc,
so that  the 21 cm signals possess slowly-varying components at $\sim 1$ MHz scales in frequency space due to a specific distribution of HII bubble sizes. For example, Mesinger \& Furlanetto (2007) predicted that  at $z=8$ the diameter distribution of the HII bubbles is peaked at ${D}_{\rm bubble}^{\rm peak}|_{z=8}\simeq 16$ Mpc (corresponding to a frequency span of $\Delta \nu_{\rm bubble}^{\rm peak}\simeq 0.9$ MHz), along with a significant tail out to $D_{\rm bubble}^{\rm max}|_{z=8}\simeq 60$ Mpc ($\Delta \nu_{\rm bubble}^{\rm max}\simeq 3.4$ MHz).
Apparently, the non-negligible slowly-varying 21 cm components contributed by these giant bubbles should have been entangled with the foreground in frequency space, and most of them are bound to be abandoned in foreground subtraction,
especially when a single-narrow-frequency segment (e.g., $\Delta\nu_{\rm segment}\simeq2$ MHz as used in Wang et al. 2006) is  adopted in the quadratic polynomial fittings.
In fact, there is a significant power loss on  Mpc scales as will be shown in \S3.2.1.

In order to address the problems raised above, we attempt to re-examine and improve the polynomial fitting algorithm by applying a more complex foreground emission model that contains  detailed emulations of both spatial and spectral features of the real sources (see Wang et al. 2010 for details, which followed Snellen et al. (2000), Giardino et al. (2002), Finkbeiner (2003), Wilman et al. (2008), and references therein; W2010 hereafter), meanwhile employing a multi-narrow-frequency segment quadratic polynomial fitting technique to remove the foreground emission components in frequency space. We find that our new
algorithm can reasonably correct the systematic bias introduced by the single-narrow-segment algorithm, which may lead to a reduction of the  21 cm signal power by about 75\% for $\gtrsim 6h^{-1}$ Mpc scales and 34\% for $\gtrsim 1h^{-1}$ Mpc scales, and therefore a significant underestimate of both the HII bubble size and growth rate. 
We also prove that at $z=0.5$ our algorithm is efficient and reliable in restoring mass distribution on $\gtrsim 30h^{-1}$ Mpc scales (by contrast about $19\%$ power will be lost with  the single-segment approach), thus it can be applied to extract the 21 cm signals emitted at intermediate redshifts  to investigate the baryon acoustic oscillations (BAO), which exhibit a typical scale of $\simeq 150$ comoving Mpc (Morales \& Wyithe 2010; Ansari et al. 2012).

 In parallel to the polynomial fitting technique,
recently a series of works have been published in an attempt to develop non-parametric techniques and apply them in observing the 21 cm signals, which do not assume a specific form for the contaminating foregrounds. Using Wp
smoothing method, Harker et al. (2009b, 2010) preferentially
consider foreground models with as few inflection points
as possible, which when apply to simulated LOFAR-EoR
data compare very favorably with parametric methods. FastICA as presented in Chapman et al. (2012) accurately
recovers the 21 cm power spectra by considering the statistically independent components of the foregrounds.
With Generalized Morphological Component Analysis (GMCA), Chapman et al. (2012b) not only recovers the
power spectra to high accuracy but also recovers the simulated 21 cm signal maps exceedingly well
using wavelet decomposition.
More detailed discussion on the non-parametric techniques can be found in the works listed above, and will not be presented here because it is beyond the scope of this paper. 

Throughout the work we adopt $h=0.71$, $\Omega_{\rm M}=0.27$, $\Omega_{\Lambda}=0.73$, and $\Omega_b=0.044$.
Unless otherwise stated, all errors are  quoted at  68\% confidence level.
\section{SIMULATION OF LOW-FREQUENCY RADIO SPECTRA}

\subsection{21 cm Signals of Neutral Hydrogen from EoR}

According to Wang et al. (2006) and references therein, the three-dimensional (3D) power spectrum of the  21 cm
signals from EoR at redshift $z$ shall take the form of
\begin{equation}
P_{\rm 3D,21cm}(k,z)=(0.016~{\rm ~K})^2\frac{1}{h^2}
\Big(\frac{\Omega_{\rm b}h^2}{0.02}\Big)^2\frac{1+z}{10}\frac{0.3}
{\Omega_{\rm M}}\left\{\left[1-x_{\rm e}(z)\right]^2+b^2(z)e^{-k^2R^2(z)}x_{\rm
e}^{2}(z)\right\}P_{\rm 3D,matter}(k,z),
\end{equation}
where $P_{\rm 3D,matter}(k,z)$ is the 3D matter power spectrum  at redshift $z$, $x_{\rm e}(z)$
is the average ionization fraction, $b(z)$ is the mean halo bias (the mean ratio of the mass over-densities of halo to field weighted by different halos),
and $R(z)=100\left[1-x_e(z)\right]^{-1/3}$ kpc is the mean radius of the ionized patches in $\rm HII$ regions.
To calculate $P_{\rm 3D,21cm}(k,z)$, we determine $b(z)$ and $x_{\rm e}(z)$ by adopting the parameters used in the fiducial reionization model of Santos et al. (2003, 2005).
At any given redshift $z_0$, we consider a redshift segment defined between $z_0$ and $z_0+\Delta z_{\rm segment}$ (corresponding to reference frequency $\nu_0$ by $\nu_0-\Delta\nu_{\rm segment}<\nu\le\nu_0$ since $\nu= {1420.4/(1+z)~{\rm MHz}}$ for the redshifted 21 cm signals), where $\Delta z_{\rm segment}\ll1$ so that $P_{\rm 3D,21cm}(k,z_0)$  is roughly uniform and isotropic within the segment $\Delta z_{\rm segment}$ or $\Delta \nu_{\rm segment}$ (i.e., the impact of evolution effects can be ignored across $\Delta z_{\rm segment}$). In this narrow redshift  range we are allowed to convert the derived $P_{\rm 3D,21cm}(k,z_0)$
into the one-dimensional (1D) power spectrum using the integration formula
\begin{equation}
P_{\rm 1D,21cm}(k)=\int_{k}^{+\infty}P_{\rm 3D,21cm}(k',z_0)k'dk'
\end{equation}
(Peacock 1999). The 1D power spectra  are shown in Figure 1 for some typical redshifts.
The spectrum (i.e., line-of-sight distribution) of the redshifted 21 cm
signals in  the segment $\nu_0-\Delta\nu_{\rm segment}<\nu\le\nu_0$ is then calculated through the inverse
Fourier transform
\begin{equation}
T_{\rm 21cm}(x_{n})=\frac{\sqrt{2\pi}}{L}\sum_{q=0}^{N-1}\bigg[A_{q}(P_{\rm 1D,21cm}(k,z_0))\cos\Big(\frac{2\pi
q}{L}x_{n}\Big)+B_{q}(P_{\rm 1D,21cm}(k,z_0))\sin\Big(\frac{2\pi
q}{L}x_{n}\Big)\bigg],
\end{equation}
where $L$ is the comoving space scale corresponding to $\Delta z_{\rm segment}$, terms
$A_{q}(P_{\rm 1D,21cm}(k,z_0))$ and  $B_{q}(P_{\rm 1D,21cm}(k,z_0))$ are independent and random parameters sharing the same Gaussian distribution ${\cal N}(0,\sqrt{{P_{\rm 1D,21cm}(k,z_{0})}/{2}})$, $k={2\pi q}/{L}$ is the wave number, $x_{n}$ ($n=1,\ldots,N$) is the line-of-sight location related to  frequency $\nu_{n}$ under  linear approximation in  frequency space, and $N$ is the number of frequency channels for the segment chosen  large enough so that most information of  $P_{\rm 1D,21cm}(k,z)$ is included within the range  $2\pi/L \le k \le 2\pi N/L$.
We adopt a segment width of $\Delta\nu_{\rm segment}=2~{\rm MHz}$ and $N=50$, therefore the frequency resolution (i.e., the width of each frequency channel) is $d \nu= \Delta \nu_{\rm segment} / N= 40~{\rm kHz}$, which is a sufficiently high frequency resolution in order to reconstruct the local power spectrum for the give $\Delta z_{\rm segment}$ at $z$.

\subsection{Foreground Emission Components and Instrumental Noise}
\subsubsection {Foreground Components Simulation}
The foreground emission components that contaminate the 21 cm signals have been continuously discussed and simulated in  detail in many works (e.g., Jeli{\'c} et al. 2008; Gleser et al. 2008; Bowman et al. 2009; Jeli{\'c} et al. 2010).
In our previous work (W2010), we fully took into account the effects of random variations of model parameters in the ranges allowed by the observations in Monte-Carlo simulations to model the emissions from our Galaxy, galaxy clusters, and discrete sources (i.e., star-forming galaxies, radio-quiet AGNs, and radio-loud AGNs), and constructed the $50-200$ MHz radio sky maps with a frequency resolution of 40 kHz, following the works of Snellen et al. (2000), Giardino et al. (2002), Finkbeiner (2003), Wilman et al. (2008), and reference therein.  All the simulated sky maps are plotted in a circular field of view (FOV) with a radius of  $5^{\circ}$  centered at the north celestial pole, as currently being observed with the 21CMA array, and the obtained images are pixelated with grids of $1024\times1024$ pixels, each image pixel covering approximately a $0.6^{\prime} \times 0.6^{\prime}$ patch. 
We adopt the same model to create the contaminating foreground emission components used in this paper.  To compare with Wang et al. (2006), the size of each sky pixel is fixed to  12 arcmin$^2$, which is  no smaller than the area covered by the point spread function of  any operating or upcoming facility designed to detect EoR signals. We note that the foreground simulations in Wang et al. (2006) did not include discrete sources brighter than 0.1 mJy at 150 MHz, because the authors considered that all of them could be identified and removed safely before the step of signal separation. In our work, however, we assume a conservative flux cut of 10 mJy at 150 MHz, since technically it is more realistic with current instrumentations (Liu et al. 2009; Pindor et al. 2011).  The brightness temperatures of the simulated foreground at 150 MHz are listed in Table 1.

\subsubsection{Instrument and Noise}
 The 21CMA array is a low-frequency radio interferometer  array constructed in a remote area of Xinjiang, China. The array consists of 81 telescopes that are distributed in a ``T'' shape with the north-south and east-west baselines being 6 km and 4 km, respectively (Figure 2), and each telescope is an assembly of 127 logarithmic periodic antennas. 
The 21CMA array is designed to continuously observe the same patch of the sky around NCP (i.e., 24 hrs a day).
The integration uv coverage of 24 hrs is shown in Figure 3.
The thermal noise of 21CMA can be expressed in terms of brightness temperature as
\begin{equation}
\sigma_{\rm noise}=\frac {T_{\rm sys} \lambda^2} {A_{\rm eff} \Omega \sqrt{ n (n-1) \tau d \nu }}
\end{equation}
(e.g., Thompson et al. 2001; W2010), where $T_{\rm sys}$ is the system temperature,  $\lambda$
is the wave length, $A_{\rm eff}$ is the
effective area of single radio telescope, $\Omega$ is the beam solid angle,  $\tau$ is the effective integral time,  and $n$ is the number of single radio telescopes.
For 21CMA, we have $T_{\rm sys}=300~ {\rm K}$, $A_{\rm eff}=218 ~{\rm m^2}$,  and $n=81$ (Table 2), thus  
\begin{equation}
\sigma_{\rm noise}=425 {\rm~ mK}\left(\frac{\lambda}{2{~\rm m}}\right)^2\left(\frac{1 \rm{~ MHz}}{d \nu}\right)^{1/2}\left(\frac{30 \rm{~days}}{\tau}\right)^{1/2}\left(\frac{1 \times 10^{-7}\rm{~sr}}{\Omega}\right).
\end{equation}
 Supposing a typical 21CMA survey of $d \nu=40 ~{\rm kHz}$, $\tau= 1$ yr, and $\Omega=1\times 10^{-6}$ sr (i.e., 12 arcmin$^2$), the noise is $\sigma_{\rm noise}|_{\nu=157.8{\rm MHz}}=60$ mK at $\lambda=1.9$ m (corresponding to $\nu=157.8$ MHz) where the 21 cm signals emitted at $z=8$ are expected to appear. Under the same observation conditions the noise at any frequency is given as $\sigma_{\rm noise}(\nu)= 60~{\rm mK} ~(\frac {\nu} {157.8~{\rm MHz}})^{-2}$. Following McQuinn et al. (2006), Chapman et al. (2012), and Zaroubi et al. (2012), the noise at each frequency $\nu$ is simulated as a Gaussian random field in the 21CMA uv plane, and then transformed to the image plane to obtain the noise  $T_{\rm noise}(\nu)$, whose  root-mean-square (rms hereafter) value is normalized to $\sigma_{\rm noise}(\nu)$.

\section{SIGNAL SEPARATION}

\subsection{ Single-Segment Fits to Pure Complex Foreground}
To extract the foreground components, n-order polynomial models are usually applied in the frequency space, e.g., Wang et al. (2006) performed  a logarithmic form
\begin{equation}
\log {T}_{\rm fore}^{\,\prime}(\nu) = a_0 + a_1 \log \nu + a_2 \log^2 \nu
\end{equation}
to approximate the integral foreground in a single segment with a width of $\Delta\nu_{\rm segment} \simeq 2$ MHz for $z_0=8$, where $a_i ~(i=0,~1,~2)$ are the parameters all obtained by applying the least squares fitting method. To determine if this approximation is appropriate when complex foreground is introduced, and if it can be applied to a wider frequency segment of a width $\sim 10$ MHz as will be discussed in \S3.2 and \S4.1,
we fit the pure simulated integral foreground in the logarithmic frequency space using linear polynomial ($\log {T}_{\rm fore}^{\,\prime}(\nu) = a_0 + a_1 \log \nu$), quadratic polynomial (Eq. (6)), and cubic  polynomial ($\log {T}_{\rm fore}^{\,\prime}(\nu) = a_0 + a_1 \log \nu + a_2 \log^2 \nu+a_3 \log^3 \nu$), respectively. Each of the three polynomial fittings is applied to 1000 randomly selected sky pixels on the simulated foreground map with different segment widths (i.e.,  $\Delta\nu_{\rm segment}=2i$ MHz for $z_0=8$, where $i=1,~ 2,~ \cdots,~ 40$) with central frequency located at $\nu_0=1420.4~{\rm MHz}/(1+z_0)$. As shown in Figure 4,  
 the intensity residuals of linear polynomial fittings can reach up to $\sim 10$ mK in many cases, which is comparable with the intensity of the 21 cm signals,  whereas the residuals of quadratic polynomial  fittings ($\lesssim 1$ mK) are  small enough to be negligible compared to the intensities of both the instrumental noises and the 21 cm signals. 
Since the quadratic polynomial fitting is good enough to approximate the complex foreground for a large range of $\Delta\nu_{\rm segment}$, we decide that the cubic one with even smaller residuals is unnecessary in our work.

In Figure 4, it is also shown that the residuals, although statistically small enough to make the quadratic polynomial fittings acceptable when $\Delta \nu_{\rm segment}<100$ MHz, increase as the segment broadens. 
We ascribe this to the fact that our foreground model  (for details see \S2.2.1 and W2010) is complex, thus a broader segment  means  that more complicated behaviors of the foreground have been involved to cause larger uncertainties. We note that the clustering of the extragalactic discrete sources and galaxy clusters, which has been taken into account in our sky simulation, does not lead to  significant residuals to overwhelm the 21 cm signals in the separation process. 
Apparently, there may exist other ways to exclude the complex foreground, e.g., the non-parametric technique (e.g., Harker et al. 2009b; Chapman et al. 2012), but it is beyond the scope of this paper to test and discuss them. 

\subsection{Separation at Noise Level of 60 mK}

The ideal low noise level as applied in some foregoing works (e.g., Wang et al. 2006) might not be achieved until the next generation facilities have been built. For currently operating facilities, such as 21CMA and LOFAR, the noise will dominate the 21 cm signals from the EoR, so that direct restoration of the 21 cm signals is precluded (e.g., McQuinn et al. 2006; Gleser et al. 2008; Bowman et al. 2009; Harker et al. 2009b; Chapman et al. 2012). To study this case, here we assume a noise level of $\sigma_{\rm noise}|_{z=8}=60$ mK, which is achievable with a one-year observation of 21CMA (\S2.2.2), and investigate how to restore the 21 cm signals.

\subsubsection{ Separation in  Single-Narrow-Frequency Segment}

As a first step, we apply the separation approach on a single segment with a narrow width of $\Delta\nu_{\rm segment}=2$ MHz (the channel width $d \nu=\Delta\nu_{\rm segment}/N=2 {~\rm MHz}/50=40$ kHz), to examine if the information on the fluctuations  of the 21 cm signals on Mpc scales is lost when our complex foreground model is applied. The segment width of $2$ MHz is adopted to guarantee that the power spectrum of the 21 cm signals is roughly uniform within the corresponding redshift segment ($\Delta z_{\rm segment}\simeq 0.1$) and $P_{\rm 1D,21cm}(z)$ can be measured accurately.

We fit the total (foreground + noise + 21 cm signals) brightness-temperature spectrum ${T}{_{\rm tot}}(\nu)$  with Equation (6)
to determine the best-fit foreground ${T}_{\rm fore}^{\,\prime}(\nu)$, and derive the residuals (raw spectrum hereafter)
 $T^{\rm raw}_{\rm 21cm}(\nu)\equiv{T}{_{\rm tot}}(\nu)-{T}_{\rm fore}^{\,\prime}(\nu)$, which contain the 21 cm signals and a dominating noise component (i.e., $T^{\rm raw}_{\rm 21cm}(\nu)=T_{\rm 21cm}^{\,\prime}(\nu)+T_{\rm noise}^{\,\prime}(\nu)$).
 Then, we transform the raw spectrum into the corresponding power spectrum. 
By subtracting the theoretically predicted noise component from the averaged power spectrum for a sufficiently large number of sky pixels, we may obtain a mean restored power spectrum for the 21 cm signals (e.g., Harker et al. 2010; Chapman et al. 2012).
When comparing the results with the input 1D power spectra created in the simulations (\S2.1; Fig. 1), we find that, when 10000 sky pixels (12 arcmin$^2$ for each) that cover about 1/3 FOV of the 21CMA (about 33 deg$^2$) are in use, the power spectrum of the 21 cm signals can be well recovered on small scales of $\lesssim 3h^{-1}~\rm Mpc$, whereas on larger scales the deviation is huge (reduced $\chi^2=5196$; Figure 5a). We measure the relative Mpc-scale power restoration in terms of the ratio of restored to simulated powers as $w|_{k<1h/{\rm Mpc}}\equiv \frac {P_{\rm 21cm}^{\prime}} {P_{\rm 21cm}}|_{k<1h/{\rm Mpc}}=\frac{\int_{2\pi/L}^{1} P_{\rm 1D,21cm}^{\prime}(k) dk} {\int_{2\pi/L}^{1} P_{\rm 1D,21cm}(k) dk}$, where $P_{\rm 21cm}|_{k<1h/{\rm Mpc}}=\int_{2\pi/L}^{1} P_{1D,{\rm 21cm}}(k) dk$ is the Mpc-scale power of the input 21 cm signals. We find  that the mean value is $\bar{w}|_{k<1h/{\rm Mpc}}=0.25\pm0.04$ for the total 10000 cases, which indicates that the use of single-narrow-segment quadratic polynomial fitting technique may cause an average power loss on Mpc scales (i.e., $\bar{w}_{\rm loss}|_{k<1h/{\rm Mpc}}=1-\bar{w}|_{k<1h/{\rm Mpc}}$) of about $75\%$ in the redshifted 21 cm signals. Since the existence of such Mpc-scale structures is expected to be quite common in late phases of EoR (e.g., Furlanetto et al. 2004b; Petrovic \& Oh 2011), part of the 21 cm signals is bound to be abandoned along with the foreground if a separation algorithm based on single-narrow-segment quadratic polynomial fittings is applied.

We exclude the possibility that such a  Mpc-scale bias is caused by the complex foreground employed in this work, simply by replacing our foreground  with that of Wang et al. (2006)'s and obtaining similar results.  An enlargement of $\Delta\nu_{\rm segment}$ to the order of 10 MHz has been suggested in some previous work (e.g., Harker et al. 2009b; Petrovic \& Oh 2011) to mitigate the Mpc-scale bias. However, as will be discussed in \S3.2.3 and \S4, we do not suggest this method because it will introduce too many auxiliary frequency channels in calculation, a relatively larger bias of $\bar{w}|_{k<1h/{\rm Mpc}}$, and a relatively poorer determination of the physical conditions of EoR when compared with the three-narrow-segment separation approach. 

\subsubsection{ Separation in Three-Narrow-Frequency Segment}

 In the following separations, we attempt to introduce more segments and frequency channels into the spectral fittings as auxiliaries to determine the foreground more accurately. 
Inspired by the fact that the complex foreground can be well approximated with quadratic polynomial expansion in  frequency ranges much wider than the largest scales of the 21 cm signals (\S3.1),
 we propose to apply a multi-narrow-segment (hereafter  multi-segment) quadratic polynomial fitting technique as an attempt to solve the problems raised in \S3.2.1. The frequency resolution and the number of channels in each segment are $d\nu=40$ kHz and $N=50$, respectively, and the span of the segments is of the order of 10 MHz.  We carry out the foreground fittings by using the same quadratic polynomial as formulated in Equation (6) to approximate the low-frequency foreground  simultaneously in three narrow segments, which are separated far away enough  to guarantee that the 21 cm signals in different segments come from different ionized  bubbles. Taking the redshift $z_0=8$ (i.e., reference frequency $\nu_0=1420.4~{\rm MHz}/(1+z_0)=157.8$ MHz) as an example, 
we define the three segments in such a way that each of them possesses the same bandwidth of $\Delta\nu_{\rm segment}=2$ MHz as in \S3.2.1, and they are separated by a gap measured either in redshift (i.e., a segment combination of $z_{\rm segment}=\{z_0,~z_1,~z_2\}$. Here $z_0$ is the reference point for segment one, and $z_1=z_0+\Delta z_{\rm gap}$ and $z_2=z_0-\Delta z_{\rm gap}$ for segment two and three, where $\Delta z_{\rm gap}= 0.5,~ 1.0,~ 1.5$, respectively) or in frequency (i.e., a segment combination of  $\nu_{\rm segment}=\{\nu_0,~\nu_1,~\nu_2\}$. Here $\nu_0$ is the reference point for segment one, and $\nu_1=\nu_0+\Delta \nu_{\rm gap}$ and $\nu_2=\nu_0-\Delta \nu_{\rm gap}$ for segment two and three,  where $\Delta \nu_{\rm gap}= 10, ~ 20, ~ 30$ MHz, respectively).
We use the same method as described in \S3.2.1 to restore the 21 cm signals, and evaluate the quality of the results in $\nu_0-2~{\rm MHz}<\nu\leq \nu_0$ using the reduced $\chi^2$ .
In general  the results are significantly better than those of the single-narrow-segment and there exists a very weak dependence of separation quality on the segment gap size. 
 Taking the segment combination of $z_{\rm segment}=\{8,~9,~7\}$ for example, the power spectrum of the 21 cm signals can be recovered very well with the three-narrow-segment approach (reduced $\chi^2=1.137$ and $\bar{w}|_{k<1h/{\rm Mpc}}=1.00\pm0.05$; Figure 5c),
which is significantly better than that obtained with the single-narrow-segment separation ($\chi^2=5196$ and $\bar{w}|_{k<1h/{\rm Mpc}}=0.25\pm0.04$), especially on large scales.
It is found that the use of more than three segments is redundant. It does not help improve the separation quality but lower the efficiency of calculation.

\subsubsection {Single-Segment Separations with a Larger $\Delta\nu_{\rm segment}$}

In order to reduce the systematic bias introduced by possible mis-subtraction of the Mpc-scale components in  redshifted 21 cm signals in single-narrow-segment quadratic polynomial fittings, Petrovic \& Oh (2011)  qualitatively mentioned that a larger segment width  should be adopted, especially in restoring the 21 cm signals in  late reionization stages when the HII bubble size becomes sufficiently large. The similar problem was also briefly mentioned by Wang et al. (2006),  who suggested an enlargement of $\Delta\nu_{\rm segment}$ by a factor of ten. To investigate if this is feasible we repeat the single-segment separation calculations described in \S3.2.1 for another 10000 rounds by increasing the segment width to $\Delta\nu_{\rm segment}=20$ MHz {\bf($N=\Delta\nu_{\rm segment}/d{\nu}=500$)} and examine the changes in the results.
The goodness of fittings is much better than that of fittings with single-narrow-segment fittings (Table 3), which supports the ideas of Petrovic \& Oh (2011) and Wang et al. (2006).
However, we note that in single-wide-segment fittings the goodness of both the reduced $\chi^2=1.365$ and $\bar{w}|_{k<1h/{\rm Mpc}}=0.97\pm0.05$ (Figure 5) are a bit worse when compared with those of our three-narrow-segment fittings (reduced $\chi^2=1.137$ and $\bar{w}|_{k<1h/{\rm Mpc}}=1.00\pm0.05$).
In addition, as will be shown in \S4.2, single-wide-segment quadratic polynomial separation technique cannot yield the best constraints on the physical conditions of EoR. 

We have also tried to enlarge $d\nu$ from 40 kHz to 100 kHz, which would reduce the required observing time to achieve $\sigma_{\rm noise}|_{\nu=157.8{\rm MHz}}=60$ mK. The results are shown in Figure 5(d)-(f), in which we can find our three-narrow-segment method still works well, whereas the information of smaller scales would be lost due to the sampling reduction of frequency.

\section{DISCUSSION}

\subsection{Further Comparison between Different Approaches}

The differences between the results obtained with single-segment and multi-segment separation approaches can be explained as follows.
 In the late stage of cosmic reionization (since $z\lesssim 9$ depending on models; e.g., Morales \& Wyithe 2010 for a recent review; Santos et al. 2008) most of the HII bubbles have been well developed. 
For example, by simulating the reionization process using the Lagrangian perturbation theory, Mesinger \& Furlanetto (2007) predicted that  at $z=8$ the bubble size ${D}_{\rm bubble}$ (corresponding to a frequency span of $\Delta\nu_{\rm bubble}$) should range from ${D}_{\rm bubble}^{\rm min}|_{z=8}\simeq 1$ Mpc ($\Delta\nu_{\rm bubble}^{\rm min}|_{z=8}\simeq 0.05 $ MHz) to ${D}_{\rm bubble}^{\rm max}|_{z=8}\simeq 60$ Mpc ($\Delta\nu_{\rm bubble}^{\rm max}|_{z=8}\simeq 3.4$ MHz) with a clear peak at ${D}_{\rm bubble}^{\rm peak}|_{z=8}\simeq 16$ Mpc ($\Delta\nu_{\rm bubble}^{\rm peak}|_{z=8}\simeq 0.9$ MHz) in number distribution.
This means that the 21 cm signals can possess slowly-varying components in frequency space on $\gtrsim 1$ MHz scales. 
In single-narrow-segment (e.g., $\Delta\nu_{\rm segment}=2$ MHz) separations,
the HII bubble size $\Delta\nu_{\rm bubble}$ is comparable to or even larger than the chosen $\Delta \nu_{\rm segment}$. 
This makes it inevitable that a significant part of the Mpc-scale 21 cm components is treated as a part of foreground, which is smooth on $\sim 100$ MHz scales, and then mis-subtracted (Fig. 6). In single-segment separations with a larger $\Delta\nu_{\rm segment}$, as the segment width increases to $\Delta \nu_{\rm segment}\gg\Delta\nu_{\rm bubble}$, the contributions of more than one HII bubbles located along line of sight are involved in the spectra to be fitted. Since the bubble and foreground components vary on quite different scales within the segment, the problem of mis-subtraction can be significantly mitigated, except that the power restorations for Mpc-scale components are not fully constrained (Table 3),
mostly due to the appearance of giant bubbles that contribute smooth components on  tens of Mpc scales.
  In three-narrow-segment separation fittings with narrow $\Delta \nu_{\rm segment}$,   the  Mpc-scale 21 cm components  in different segments come from different bubbles, i.e., they do not correlate with each other whereas the foreground components in the three segments do. Therefore the foreground and 21 cm signals are best determined, respectively.

This can also be clearly seen when the noise is sufficiently low (e.g., $\sigma_{\rm noise}|_{z=8}=1$ mK), although the realistic condition (e.g., $\sigma_{\rm noise}|_{z=8}=60$ mK) follows the same rule. Taking $\sigma_{\rm noise}|_{z=8}=1$ mK as an example, we repeat the separation using single-narrow-segment, three-narrow-segment, and single-wide-segment separation approaches with 1000 rounds, respectively.  
 We evaluate the result of each round in a quantitative way by calculating the  relative root-mean-square error (RRMSE hereafter) as follows,
$r_{\rm fit}\equiv\frac {\sqrt{\frac {1} {N} \sum_{i=1}^{N} {\left[ T^{\rm raw}_{\rm 21cm}(\nu_i)- {T}_{\rm 21cm}(\nu_i)\right]^2}}} { {|T_{\rm 21cm}(\nu)|}_{\rm MAX}}$.
With the single-narrow-segment approach, only $25\%$ of the 1000 cases can yield acceptable ($r_{\rm fit}\leq 0.1$; Figure 7 and see an example in Figure 8) results. We find that failed  ($r_{\rm fit}>0.1$; Figure 7 and see an example in Figure 9) restorations tend to occur when the 21 cm signals possess significant large-scale ($k\lesssim 1{h ~\rm Mpc^{-1}}$) components  along line of sight, which contribute a non-negligible part to the simulated 21 cm spectra (Figure 10). The best restoration of the 21 cm signals can be obtained with the three-narrow-segment fitting technique (Figures 7 and 9; Table 4), and in single-wide-segment fittings the scatter of power restoration on  $\gtrsim 6h^{-1}$ Mpc  scales is larger when compared with our three-narrow-segment fittings (Table 4).
We also show the result of the two-narrow-segment approach, which is similar to  three-narrow-segment approach but has one less segment. In general  the results of two-narrow-segment approach are better than those of the single-narrow-segment separations, but
the goodness of separation is more or less sensitive to the choice of $\Delta z_{\rm gap}$ or $\Delta \nu_{\rm gap}$ (Figure 7). 

We also have applied above separation approaches to higher redshifts of $z_0=$9, 10, 12, and 14, which correspond to reference frequencies of $\nu_0=$142.0 MHz, 129.1 MHz, 109.3 MHz, and 94.7 MHz for redshifted 21 cm signals, respectively (Table 5).
At these  redshifts, where the cosmic reionization model adopted in this work (Santos et al. 2003, 2005; \S2.1) predicts that the HII bubbles are growing in their earlier stage and 
less extended, the separation quality of our three-narrow-segment approach is still the best, although statistically the separation quality of the single-wide-segment approach is almost equally good.

\subsection {Constraints on EoR Physics}

 In fiducial reionization models such as the one adopted in this work (Eq. (1)), the power of the redshifted 21 cm signals is mainly determined by the mean halo bias $b$ and the average ionization fraction $x_e$ at a given redshift (Fig. 11), whereas nowadays both $b$ and $x_e$ are poorly constrained and suffer large uncertainties among different theoretical  reionization models (Santos et al. 2003). 
In future observations, well restored 21 cm powers will greatly help derive $b$ and $x_e$, and then deduce HII bubble size and growth rate. Since powers integrated in different scale ranges follow different contour lines on the $x_e-b$ diagram (e.g., $w|_{k<1 h/{\rm Mpc}}$ and $w|_{k<5 h/{\rm Mpc}}$ on Fig. 11), it is feasible to constrain $b$ and $x_e$ simultaneously using the overlapping area of any two of these contours (Fig. 12), the accuracy of which depends on how accurate we can restore the corresponding 21 cm powers. In Figure 12 and Table 3 we clearly see that, for an ideal noise condition (1 mK), the tightest constraints  ($b=3.50^{+0.25}_{-0.10}$ and $x_e=0.84^{+0.05}_{-0.07}$) are obtained using the powers restored with our three-narrow-segment separation approach, when compared with the input model values ($b=3.50$ and  $x_e=0.84$; Santos et al. 2003). When using the single-wide-segment approach, the obtained $b=3.50^{+10.5}_{-0.20}$ and $x_e=0.83^{+0.13}_{-0.65}$ are also unbiased but their errors are larger by about one order of magnitude compared with three-narrow-segment approach. When using the single-narrow-segment approach  both of the constraints ($2.85\leq b \leq 14$ and $0.14 \leq x_e \leq 0.98$) are very poor.
For a realistic noise condition (60 mK), the  three-narrow-segment separation approach still works well ($b=3.55^{+0.75}_{-0.15}$ and $x_e=0.83^{+0.07}_{-0.19}$; Fig. 13 and Table 3) whereas the single-wide-segment approach introduces larger errors in the result ($b=3.75^{+1.65}_{-0.35}$ and $x_e=0.76^{+0.12}_{-0.26}$).
In such a sense we conclude that the three-narrow-segment quadratic polynomial fitting technique is the best approach to restore the 21 cm signals.

\subsection{Detection of BAO signals at Intermediate Redshifts with Three-Segment Separation Approach}
We have also studied whether or not our three-segment separation technique can be applied to intermediate frequency bands to detect the characteristic BAO signals (Ansari et al. 2012), which can be used as a sensitive probe to investigate the nature of dark matter and to constrain cosmological models (e.g., Bharadwaj et al. 2009 for a review; Visbal et al. 2009; Bagla et al. 2010). 
We apply the single-narrow-segment, single-wide-segment and three-narrow-segment approaches to the redshift range of $0.35-0.65$, which corresponds to $861$ MHz$-1052$ MHz in detection.  The power spectrum of the brightness temperature of the 21 cm signals is assumed to follow that of the matter density field
\begin{equation}
P_{\rm 3D,21cm}(k,z)=\bar{T}^2_{\rm 21cm}(z)b^2(z)P_{\rm 3D,matter}(k,z)
\end{equation}
(Seo et al. 2010), where $\bar{T}_{\rm 21cm}(z)$ is the brightness temperature of the redshifted 21 cm
signals averaged at $z$, and $b(z)$ is the mean bias. According to Barkana \& Loeb (2007), Pritchard
\& Loeb (2008), and Chang et al. (2008), $b(z)=1$ after the cosmic reionization, and $\bar{T}_{\rm 21cm}(z)$ is estimated to be
\begin{equation}
\bar{T}_{\rm 21cm}(z)=\frac {188x_{\rm HI}(z)\Omega_{\rm H,0}h(1+z)^2} {\sqrt{\Omega_{\rm M}(1+z)^3+1-\Omega_{\rm M}}}~{\rm mK},
\end{equation}
 where $x_{\rm HI}(z)$ is the neutral hydrogen fraction at $z$,  and $\Omega_{\rm H,0}$ is the ratio
of the total hydrogen mass density to the critical density at $z = 0$. In literature (e.g., Zwaan et al. 2005; Prochaska et al.
2005; Rao et al. 2006) a conservative estimation  $x_{\rm HI}(z)\Omega_{\rm H,0}= 0.00037$  is often assumed for the neutral hydrogen fraction at intermediate redshifts.

As a simple test, we carry out both single-segment and three-segment separations ($\Delta z_{\rm gap}=0.15,~ 0.20,~ 0.25$ or $\Delta \nu_{\rm gap}=100,~ 150,~ 200$ MHz) at $z_0=0.5$ with a channel width  $d\nu = 0.1$ MHz, a channel number  $N = 500$ (i.e., a segment width $\Delta\nu_{\rm segment}=Nd\nu=50$ MHz), and $\sigma_{\rm noise}|_{z=0.5}$=20 mK. As shown in Figure 14 and Table 6, the  three-segment separation approach (e.g., $\bar{r}_{\rm fit}=0.016\pm0.004$ for $z_{\rm segment}=\{0.5,~0.7,~0.3\}$) yields better results than the single-segment separation approach ($\bar{r}_{\rm fit}=0.036\pm0.015$).
The power loss in the reconstruction for the $\gtrsim 30h^{-1}$ Mpc scale structures is $\bar{w}_{\rm loss}|_{k<0.2h/{\rm Mpc}}=1-\bar{w}|_{k<0.2h/{\rm Mpc}}\leq 0.02$ with three-segment separation approach while $\bar{w}_{\rm loss}|_{k<0.2h/{\rm Mpc}}=0.19\pm0.16$ with single-segment separation approach.
Therefore we conclude that the  three-segment separation approach  can successfully restore the 21 cm signals from
intermediate redshifts better than the single-segment approach, showing its potential to be employed in detecting the BAO signals that typically possess characteristic $\simeq 150$ comoving Mpc scales. 
More details on the simulation of the BAO signals and its separation from the foreground will be presented in the next paper since it is beyond the scope of this work.

\section{SUMMARY}

In this work we construct a complex foreground model and compare between different separation approaches by carrying out statistical analysis of the restored 21 cm spectra and corresponding power spectra, as well as their constraints on mean halo bias $b$ and average ionization fraction $x_e$.
 At $z=8$ the best restoration of 21 cm signals and the tightest determination of $b$ and $x_e$ can be obtained with the three-narrow-segment quadratic polynomial fitting technique proposed in this work. 
We illustrate that the new technique also works well at redshifts higher than 8, and is a potentially useful tool to probe the BAO at intermediate redshifts.

\begin{acknowledgements}

We sincerely thank the referee for his/her careful review and
valuable comments on our work, which help polish and improve the
manuscript.
 This work was supported by the Ministry of Science and Technology of the People's Republic of China (973 Program; Grant Nos. 2009CB824900 and
2009CB824904), the National Science Foundation of China (Grant Nos.
10878001, 10973010, 11125313, and 11203041), the Shanghai Science and Technology
Commission (Grant No. 11DZ2260700), and Shanghai Jiao Tong University
Innovation Fund for Postgraduates.

\end{acknowledgements}


\clearpage

\begin{table}[]
  
  \footnotesize
  \renewcommand{\thefootnote}{}
  \caption{Foreground Emission Components at 150 MHz  (spatial resolution = 12 arcmin$^2$).}
  \label{Tab:publ-works}
  
  \begin{center}
    \begin{tabular}{c|cc}
      \hline \hline
Component & Mean Value  & rms \\
& (K)& (K) \\
\hline
Galaxy synchrotron emission & $237$ & $24$ \\
Galaxy free-free emission & $0.877 $& $0.205$\\
Galaxy clusters & $0.084$ & $2.06$ \\
Extragalactic discrete sources &88 & 7937 \\
Extragalactic discrete sources  (under $S_{\rm cut}$) & 23 & 29 \\
\hline
Total foreground & 326 & 7937\\
Total foreground (with discrete sources under $S_{\rm cut}$) &261 & 38 \\

\hline  
    \end{tabular}

\end{center}

\end{table}

\begin{table}[]
  
  \footnotesize
  \renewcommand{\thefootnote}{}
  \caption{Parameters of our synthetic observations.}
  \label{Tab:publ-works}

 \begin{center}
    \begin{tabular}{c|cc}
      \hline \hline
Parameter & Value \\
\hline
Field of view & $10^{\circ}\times10^{\circ} $ \\
Total effective area $nA_{\rm eff}$ & 17658 m$^2$\\
Number of  baselines $n(n-1)/2$& $(81\times80)/2$ \\
Frequency coverage & $50-200$ MHz \\
Frequency resolution $d\nu$& 40 kHz \\
Image spatial resolution $\Omega$& $1\times 10^{-6}$ sr \\
Observing time $\tau$& 1 yr \\
$\sigma_{\rm noise}$ at 157.8 MHz & 60 mK \\
\hline  
    \end{tabular}

\end{center}

\end{table}


\begin{table}[]
  \footnotesize
  \renewcommand{\thefootnote}{}
  \caption{Comparison between different separation approaches -- cases at $z_0=8$  with $\sigma_{\rm noise}|_{z=8}$=60 mK and 1mK, and mean halo bias and average ionization fraction calculated from relative 21 cm power restoration$^a$.}
  \label{Tab:publ-works}
  \begin{center}
    \begin{tabular}{c|c|cc|cc}
      \hline \hline
      
      \multicolumn{2}{c|}{\multirowcell{2}[.0ex][c]{}}&\multicolumn{2}{c|}{Relative Power Restoration}&Mean Halo Bias&Average Ionization Fraction\\
      \multicolumn{2}{c|}{}&$\bar{w}|_{k<1h/{\rm Mpc}}$&$\bar{w}|_{k<5h/{\rm Mpc}}$&$b$&$x_e$\\
      \hline
      \multicolumn{2}{c|}{Input model}&1.00&1.00&3.50&0.84\\
      \hline

  \multirowcell{6}[.0ex][c]{$\sigma_{\rm noise}|_{z=8}$\\=60 mK$~~~$}&\multirowcell{2}[.0ex][c]{Single-narrow-segment\\($\Delta\nu_{\rm segment}=2$ MHz)}      &\multirowcell{2}[0ex][c]{$0.25\pm0.04$}&\multirowcell{2}[0ex][c]{ $0.66\pm0.02$ }&\multirowcell{2}[0ex][c]{ -- }&\multirowcell{2}[0ex][c]{ -- }\\
      &&&\\
      \cline{2-6}
      &\multirowcell{2}[0ex][c]{Single-wide-segment\\($\Delta\nu_{\rm segment}=2$ MHz)}      &\multirowcell{2}[0ex][c]{$0.97\pm0.05$}&\multirowcell{2}[0ex][c]{ $1.01\pm0.02$ }&\multirowcell{2}[0ex][c]{ $3.75^{+1.65}_{-0.35}$ }&\multirowcell{2}[0ex][c]{ $0.76^{+0.12}_{-0.26}$}\\
      &&&\\
      \cline{2-6}
      &\multirowcell{2}[0ex][c]{Three-narrow-segment\\($\Delta\nu_{\rm segment}=2$ MHz)} & \multirowcell{2}[0ex][c]{$1.00\pm0.05$}& \multirowcell{2}[0ex][c]{$1.01\pm0.02$} &\multirowcell{2}[0ex][c]{ $3.55^{+0.75}_{-0.15}$} &\multirowcell{2}[0ex][c]{ $0.83^{+0.07}_{-0.19}$}\\
      &&&\\
     
      \hline

      \multirowcell{6}[.0ex][c]{$\sigma_{\rm noise}|_{z=8}$\\=1 mK$~~~~$}&\multirowcell{2}[.0ex][c]{Single-narrow-segment\\($\Delta\nu_{\rm segment}=2$ MHz)}      &\multirowcell{2}[0ex][c]{$0.64\pm0.24$}&\multirowcell{2}[0ex][c]{ $0.83\pm0.13$ }&\multirowcell{2}[0ex][c]{$ [2.85, 14]$ }&\multirowcell{2}[0ex][c]{ $[0.14, 0.98]$ }\\
      &&&\\
      \cline{2-6}
      &\multirowcell{2}[0ex][c]{Single-wide-segment\\($\Delta\nu_{\rm segment}=20$ MHz)} & \multirowcell{2}[0ex][c]{$0.98\pm0.12$}& \multirowcell{2}[0ex][c]{$0.99\pm0.05$} &\multirowcell{2}[0ex][c]{ $3.50^{+10.5}_{-0.20}$} &\multirowcell{2}[0ex][c]{ $0.83^{+0.13}_{-0.65}$}\\
      &&&\\
      \cline{2-6}
      &\multirowcell{2}[0ex][c]{Three-narrow-segment\\($\Delta\nu_{\rm segment}=2$ MHz)} & \multirowcell{2}[0ex][c]{$1.00\pm0.02$}& \multirowcell{2}[0ex][c]{$1.00\pm0.02$} &\multirowcell{2}[0ex][c]{ $3.50^{+0.25}_{-0.10}$} & \multirowcell{2}[0ex][c]{$0.84^{+0.05}_{-0.07}$}\\
      &&&\\
      \hline

    \end{tabular}\end{center}
{ Note. $^a$ Calculations are carried out using  the fiducial reionization model of Santos et al. (2003, 2005).  }
\end{table}


\begin{table}[]

\footnotesize
\renewcommand{\thefootnote}{}
  \caption{Comparison between different separation approaches -- cases at $z_0=8$  with $\sigma_{\rm noise}|_{z=8}$=1 mK (Fig. 7; \S4.1)$^a$.}
  \label{Tab:publ-works}

  \begin{center}
\begin{tabular}{c|crcc}
  \hline \hline
  
  &\multicolumn{2}{c}{Gap Width}&$\bar{r}_{\rm fit}$&$\bar{w}|_{k<1h/{\rm Mpc}}$\\
  \hline
  
  \multirowcell{2}[0.0ex][c]{Single-narrow-segment \\($\Delta\nu_{\rm segment}= 2$ MHz)}&\multicolumn{2}{c}{\multirowcell{2}[0.0ex][c]{--}}&\multirowcell{2}[0.0ex][c]{$0.153\pm0.070$}&\multirowcell{2}[0.0ex][c]{$0.64\pm0.24$}\\
  &\multicolumn{2}{c}{}&&\\
  \hline
  
  \multirowcell{2}[0.0ex][c]
  {Single-wide-segment\\($\Delta\nu_{\rm segment}=20$ MHz)}&\multicolumn{2}{c}{\multirowcell{2}[0.0ex][c]{--}}&\multirowcell{2}[0.0ex][c]{$0.055\pm0.034$}&\multirowcell{2}[0.0ex][c]{$0.98\pm0.12$}\\
  &\multicolumn{2}{c}{}&&\\
  \hline
  
  \multirowcell{12}[0.0ex][c]
  {Two-narrow-segment\\
    ($\Delta\nu_{\rm segment}= 2$ MHz)} &
  
  \multirowcell{6}[0.0ex][c]
  {$\Delta z_{\rm gap}$}&$-1.5$~~~~&$0.116\pm0.083$&$1.09\pm0.51$\\
  &  &$-1.0$~~~~&$0.099\pm0.072$&$1.03\pm0.35$\\
  &  &$-0.5$~~~~&$0.082\pm0.054$&$0.97\pm0.23$\\
  &  &$0.5$~~~~&$0.062\pm0.037$&$0.92\pm0.16$\\
  &  &$1.0$~~~~&$0.056\pm0.034$&$0.91\pm0.14$\\
  &  &$1.5$~~~~&$0.050\pm0.028$&$0.92\pm0.13$\\
  \cline {2-5}
  &
  \multirowcell{6}[0.0ex][c]
  {$\Delta \nu_{\rm gap}$\\(MHz)}&$-30$~~~~&$0.048\pm0.026$&$0.91\pm0.12$\\
  &  &$-20$~~~~&$0.052\pm0.030$&$0.91\pm0.13$\\
  &  &$-10$~~~~&$0.062\pm0.037$&$0.91\pm0.16$\\
  &  &$10$~~~~&$0.084\pm0.059$&$0.96\pm0.22$\\
  &  &$20$~~~~&$0.098\pm0.069$&$1.02\pm0.31$\\
  &  &$30$~~~~&$0.116\pm0.086$&$1.13\pm0.84$\\
  \hline
  
  \multirowcell{6}[0.0ex][c]
  {Three-narrow-segment\\($\Delta\nu_{\rm segment}= 2$ MHz)} &
  \multirowcell{3}[0.0ex][c]
  {$\Delta z_{\rm gap}$}&$0.5$~~~~&$0.028\pm0.010$&$1.00\pm0.02$\\
  &  &$1.0$~~~~&$0.024\pm0.006$&$1.00\pm0.02$\\
  &  &$1.5$~~~~&$0.023\pm0.005$&$1.00\pm0.02$\\
  \cline {2-5}
  &
  \multirowcell{3}[0.0ex][c]
  {$\Delta \nu_{\rm gap}$\\(MHz)}&$10$~~~~&$0.027\pm0.009$&$1.00\pm0.02$\\
  &  &$20$~~~~&$0.023\pm0.006$&$1.00\pm0.02$\\
  &  &$30$~~~~&$0.022\pm0.005$&$1.00\pm0.02$\\
  \hline
\end{tabular}

\end{center}
{Note. $^a$ We run 1000 rounds of simulation and separation for each separation approach. $\bar{r}_{\rm fit}$ is the mean   relative root-mean-square error for the restoration of the 21 cm signals, and  $\bar{w}|_{k<1h/{\rm Mpc}}$ is the mean power restoration on $\gtrsim 6h^{-1}$ Mpc scales.  Errors indicate the 68\% confidence limits.}
\end{table}


\begin{table}[]
  
  \footnotesize
  \renewcommand{\thefootnote}{}
  \caption{Comparison between different separation approaches -- cases for $z_0=9,~10,~12,~14$ with $\sigma_{\rm noise}(\nu)=(\nu/{157.8 ~\rm MHz})^{-2}$ mK (\S4.1)$^a$.}
  \label{Tab:publ-works}
  
  \begin{center}\scriptsize
    \begin{tabular}{c|c|ccccc}
      \hline \hline
      $z_0$&Parameter&Single-Narrow-Segment&Single-Wide-Segment&Three-Narrow-Segment&Three-Narrow-Segment\\
      &&($\Delta\nu_{\rm segment}=2$ MHz)&($\Delta\nu_{\rm segment}=20$ MHz)&($\Delta z_{\rm gap}=1$)&($\Delta \nu_{\rm gap}=10$ MHz)\\
      \hline
      \multirow{2}{0.4cm}
      {9}&$\bar{r}_{\rm fit}$&$0.145\pm0.067$&$0.053\pm0.028$&$0.032\pm0.007$&$0.033\pm0.009$\\
      &$\bar{w}|_{k<1 h/{\rm Mpc}}$&$0.66\hspace{0.5em}\pm0.23\hspace{0.5em}$&$0.99\hspace{0.5em}\pm0.10\hspace{0.5em}$&$1.00\hspace{0.5em}\pm0.03\hspace{0.5em}$&$1.00\hspace{0.5em}\pm0.03\hspace{0.5em}$\\
      \hline
      \multirow{2}{0.4cm}
      {10}&$\bar{r}_{\rm fit}$&$0.141\pm0.062$&$0.058\pm0.025$&$0.042\pm0.010$&$0.043\pm0.009$\\
      &$\bar{w}|_{k<1 h/{\rm Mpc}}$&$0.67\hspace{0.5em}\pm0.23\hspace{0.5em}$&$0.99\hspace{0.5em}\pm0.11\hspace{0.5em}$&$1.01\hspace{0.5em}\pm0.05\hspace{0.5em}$&$1.01\hspace{0.5em}\pm0.05\hspace{0.5em}$\\
      \hline
      \multirow{2}{0.4cm}
      {12}&$\bar{r}_{\rm fit}$&$0.146\pm0.055$&$0.075\pm0.022$&$0.067\pm0.014$&$0.066\pm0.013$\\
      &$\bar{w}|_{k<1 h/{\rm Mpc}}$&$0.73\hspace{0.5em}\pm0.21\hspace{0.5em}$&$1.01\hspace{0.5em}\pm0.11\hspace{0.5em}$&$1.01\hspace{0.5em}\pm0.08\hspace{0.5em}$&$1.01\hspace{0.5em}\pm0.07\hspace{0.5em}$\\
      \hline
      \multirow{2}{0.4cm}
      {14}&$\bar{r}_{\rm fit}$&$0.157\pm0.050$&$0.099\pm0.024$&$0.094\pm0.020$&$0.093\pm0.020$\\
      &$\bar{w}|_{k<1 h/{\rm Mpc}}$&$0.75\hspace{0.5em}\pm0.22\hspace{0.5em}$&$1.02\hspace{0.5em}\pm0.13\hspace{0.5em}$&$1.02\hspace{0.5em}\pm0.11\hspace{0.5em}$&$1.02\hspace{0.5em}\pm0.11\hspace{0.5em}$\\
      \hline
      
    \end{tabular}

\end{center}
{Note. $^a$ Parameters are defined in the same way as in Table 4.}

\end{table}

\begin{table}[]
  \footnotesize
  \renewcommand{\thefootnote}{}
  \caption{Comparison between different separation approaches -- cases for $z_0=0.5$ (Fig. 14(a); \S4.3)$^a$.}
  \label{Tab:publ-works}
  \begin{center}
    \begin{tabular}{c|crcc}
      \hline \hline
      
      &\multicolumn{2}{c}{Gap Width}&$\bar{r}_{\rm fit}$&$\bar{w}|_{k<0.2h/{\rm Mpc}}$\\
      \hline
      \multirowcell{2}[0.0ex][c]{Single-segment\\($\Delta\nu_{\rm segment}= 50$ MHz)} &\multicolumn{2}{c}{\multirowcell{2}[0.0ex][c]{--}}&\multirowcell{2}[0.0ex][c]{$0.036\pm0.015$}&\multirowcell{2}[0.0ex][c]{$0.81\pm0.16$}\\
      &\multicolumn{2}{c}{}\\
      \hline
      \multirowcell{6}[0.0ex][c]
      {Three-segment\\
        ($\Delta\nu_{\rm segment}= 50$ MHz)} &
      \multirowcell{3}[0.0ex][c]
      {$\Delta z_{\rm gap}$}&0.15~~~~&$0.016\pm0.004$&$1.00\pm0.02$\\
      &  &0.20~~~~&$0.016\pm0.004$&$1.00\pm0.02$\\
      &  &0.25~~~~&$0.016\pm0.003$&$1.00\pm0.02$\\
      \cline {2-5}
      &
      \multirowcell{3}[0.0ex][c]
      {$\Delta \nu_{\rm gap}$\\(MHz)}&100~~~~&$0.016\pm0.004$&$1.00\pm0.02$\\
      &  &150~~~~&$0.015\pm0.003$&$1.00\pm0.02$\\
      &  &200~~~~&$0.015\pm0.003$&$1.00\pm0.02$\\
      \hline
    \end{tabular}\end{center}
{Note. $^a$ Parameters are defined in the same way as in Table 4, whereas the wave number range is confined as $k<0.2h~{\rm Mpc}^{-1}$ to calculate relative power restoration.}

\end{table}


\clearpage

\begin{figure}[]
     \begin{center}

\includegraphics[width=0.7\textwidth]{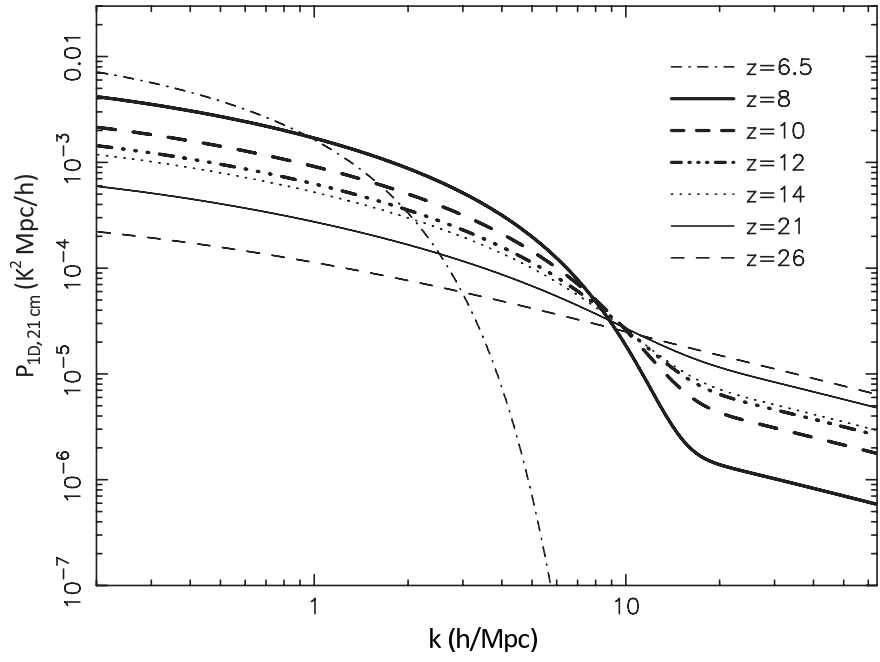}
   \caption{1D power spectra of 21 cm signals calculated at some typical redshifts  by adopting the fiducial reionization model (\S2.1; Santos et al. 2003, 2005).}
   \end{center}
\end{figure}

\begin{figure}[]
     \begin{center}

\includegraphics[width=0.7\textwidth]{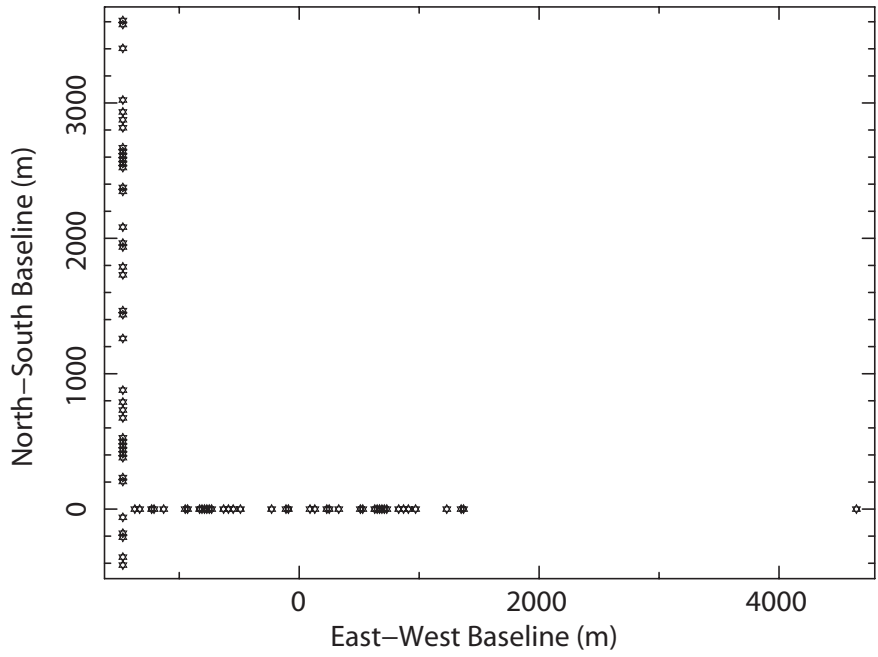}
   \caption{Telescope arrangement of the 21CMA array. Each telescope is an assembly of 127 logarithmic periodic antennas.}
   \end{center}
\end{figure}

\begin{figure}[]
     \begin{center}

\includegraphics[width=0.7\textwidth]{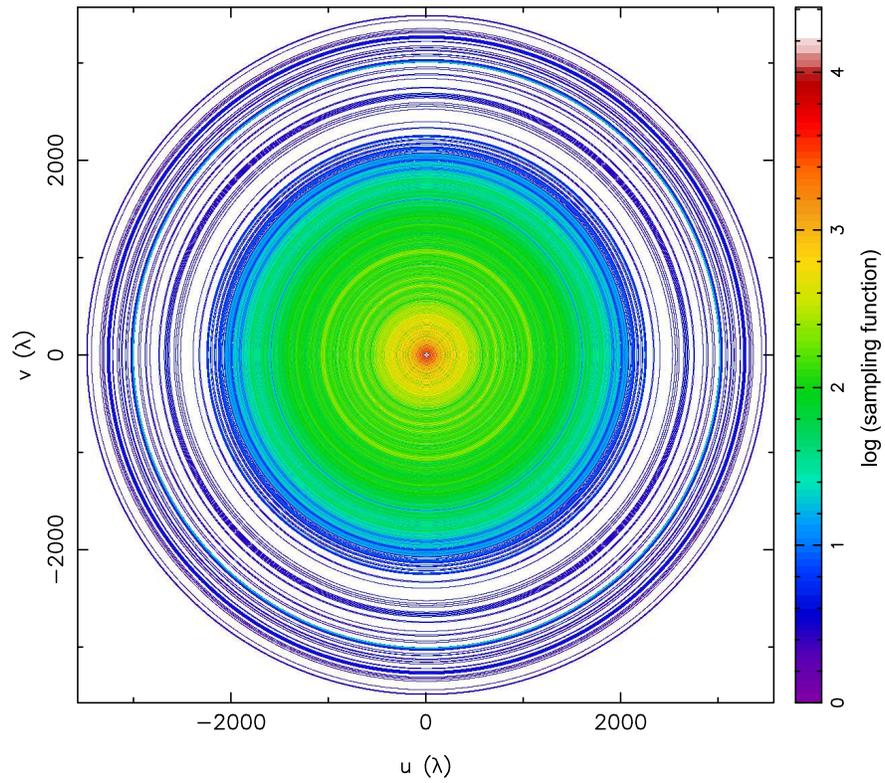}
   \caption{Density of visibility measurements in the 21CMA uv plane at
150 MHz with an integration time of 24 hours (\S2.2.2). }
   \end{center}
\end{figure}

\begin{figure}[]
    \begin{center}
\includegraphics[width=0.7\textwidth]{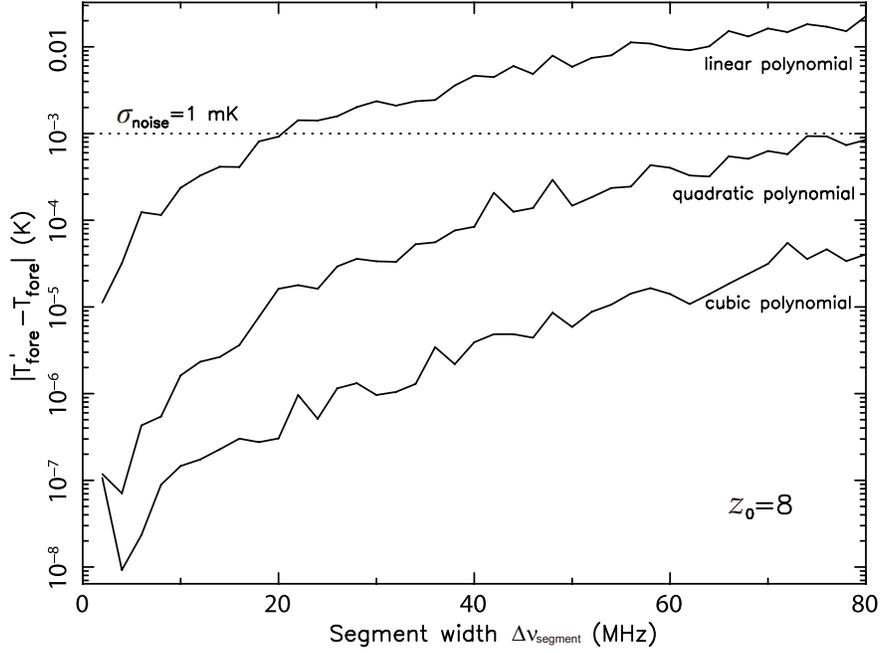}
   \caption{
Difference between restored  ($T_{\rm fore}^{\,\prime}$) and input ($T_{\rm fore}$) foregrounds calculated with single-segment separation approach by using different polynomial approximations (\S3.1). The calculations are repeated for 1000 times for each given  $\Delta\nu_{\rm segment} = 2i$ MHz ($i=1,~2,~\cdots,~40$) to determine the mean value of $|T_{\rm fore}^{\,\prime}-T_{\rm fore}|$.
}
   \end{center}
\end{figure}

\begin{figure}[]
     \begin{center}

\includegraphics[width=\textwidth]{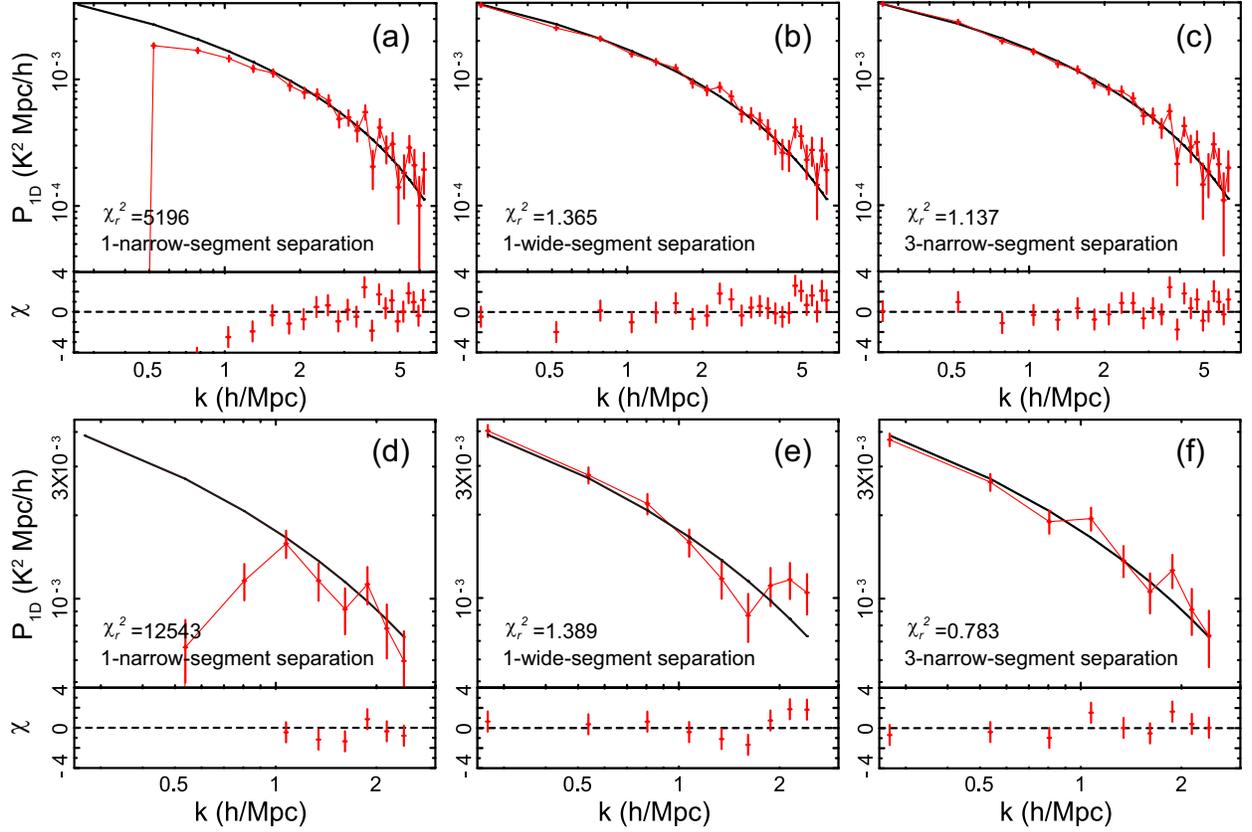}
  \caption{ Mean restored power spectra for the 21 cm signals from 10000 sky pixels with $\sigma_{\rm noise}|_{z=8}$=60 mK, as compared with the theoretical 21 cm power spectrum at $z_0=8$ (black). 
(a)-(c) are the condition with $d\nu=40$ kHz, and (d)-(f) with $d\nu=100$ kHz.
Residuals obtained in the $\chi^2$ test between the restored and theoretical 21 cm power spectra in each case are also plotted in the lower panels.}
   \end{center}
\end{figure}

\begin{figure}[]
     \begin{center}
\includegraphics[width=\textwidth]{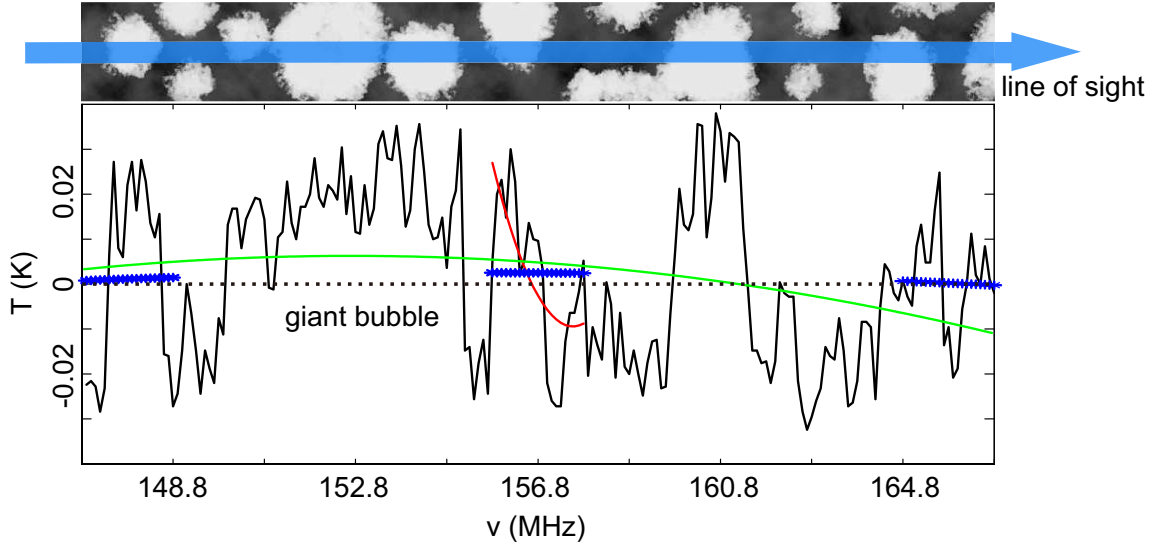}
   \caption{  
A comparison between three separation techniques discussed in \S3 and \S4. Upper panel: Sketch diagram for the spatial distribution of ionized bubbles along line of sight at $z_0=8$. Lower panel: Corresponding spectrum of the simulated 21 cm signals (black solid), along with the best-fits  (red solid: single-narrow-segment; green solid: single-wide-segment; blue star: three-narrow-segment) of $T_{\rm fore}^{\prime}-T_{\rm fore}$, which presents the Mpc-scale components of 21 cm signals that have been entangled with the foreground. Clearly the best-fit spectrum of single-narrow-segment is significantly biased by the Mpc-scale components of the bubble, and best-fit spectrum of single-wide-segment is biased by the tens of Mpc-scale components of the giant bubbles. The black dotted line marks $T_{\rm fore}^{\prime}-T_{\rm fore}=0$, which is the most ideal result for the restoration of the 21 cm signals.}
   \end{center}
\end{figure}

\begin{figure}[]
     \begin{center}

\includegraphics[width=\textwidth]{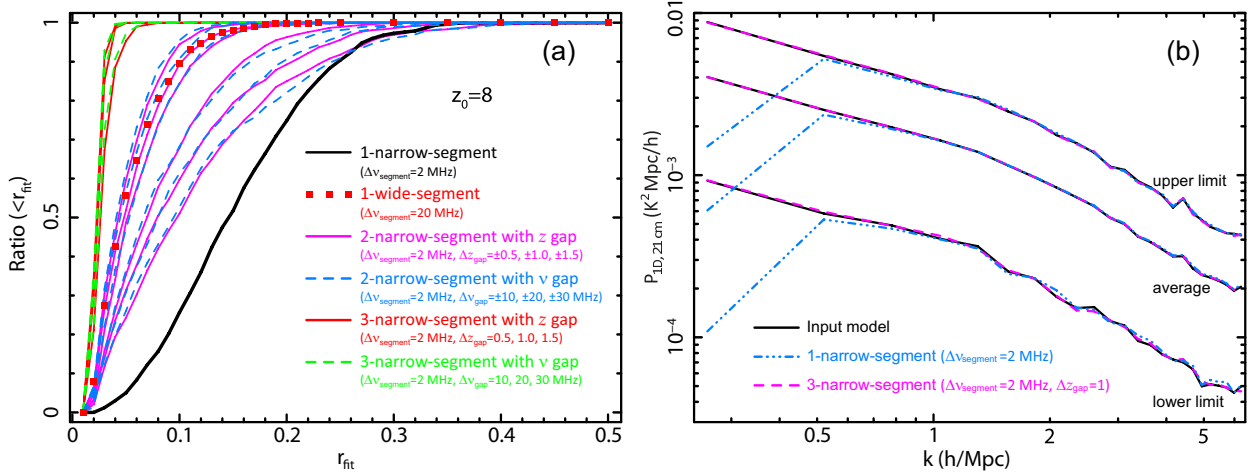}
   \caption{(a) Fractional distributions of the relative root-mean-square error $r_{\rm fit}$ for single-narrow-segment (black solid), single-wide-segment (red square), two-narrow-segment with $z$ gap (magenta solid; upper left to lower right are $\Delta z_{\rm gap}=1.5,~1.0,~\cdots,~-1.5$, respectively),  two-narrow-segment with $\nu$ gap (blue dash; upper left to lower right are $\Delta\nu_{\rm gap}=-30,~-20,~\cdots,~30$ MHz, respectively), three-narrow-segment with $z$ gap (red solid;  upper left to lower right is $\Delta z_{\rm gap}=1.5,~1.0,~0.5$, respectively), and three-narrow-segment with $\nu$ gap  (green dash; upper left to lower right are $\Delta\nu_{\rm gap}=30,~20,~10$ MHz, respectively) separation approaches, each of which is calculated based on 1000 rounds of simulation and separation with $\sigma_{\rm noise}|_{z=8}$=1 mK (Table 4; \S4.1). (b) Average power spectra of input and restored 21 cm signals and their 68\% dispersion limits. The restorations are performed with single-narrow-segment (blue) and three-narrow-segment (magenta) quadratic polynomial fitting techniques based on 1000 rounds of simulation and separation  for each.}
   \end{center}
\end{figure}

\begin{figure}[]
     \begin{center}
\includegraphics[width=\textwidth]{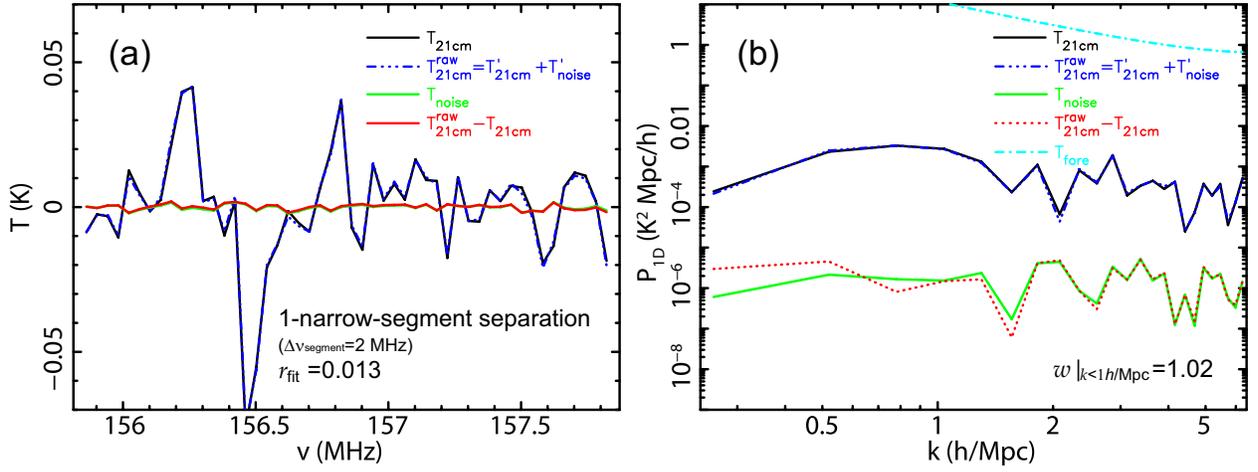}
   \caption{One case for successful separation with the single-narrow-segment quadratic polynomial fitting technique, for which the segment width is chosen to be 2 MHz, $\sigma_{\rm noise}|_{z=8}$=1 mK, and our complex foreground model is applied (\S4.1). 
(a) Spectra of input 21 cm signals $T_{\rm 21cm}$, restored 21 cm signals plus noise $T_{\rm 21cm}^{\rm raw}$, noise $T_{\rm noise}$, and the difference between  $T_{\rm 21cm}^{\rm raw}$ and $T_{\rm 21cm}$ are shown in black, blue, green and red, respectively. (b)  Corresponding power spectra for (a).  Power spectrum of the complex foreground (cyan) is also plotted for comparison.
}
   \end{center}
\end{figure}

\begin{figure}[]
     \begin{center}

\includegraphics[width=\textwidth]{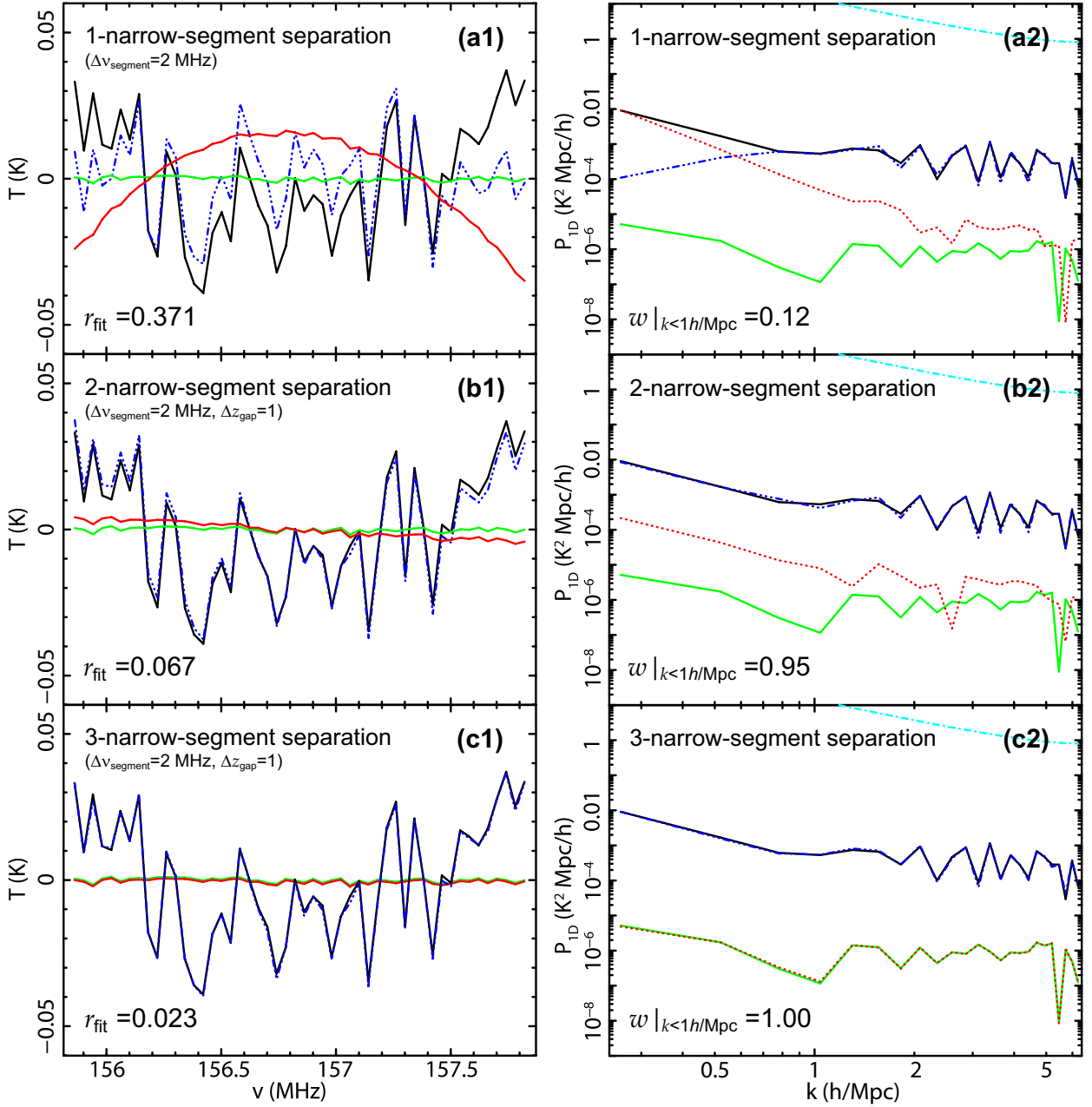}
   \caption{Same as Figure 8 but for one case of failed separation at $z_0=8$ with the single-narrow-segment approach (\S4.1). Spectra are marked in the same colors as in Figure 8.
}
   \end{center}
\end{figure}

\begin{figure}[]
     \begin{center}

\includegraphics[width=0.7\textwidth]{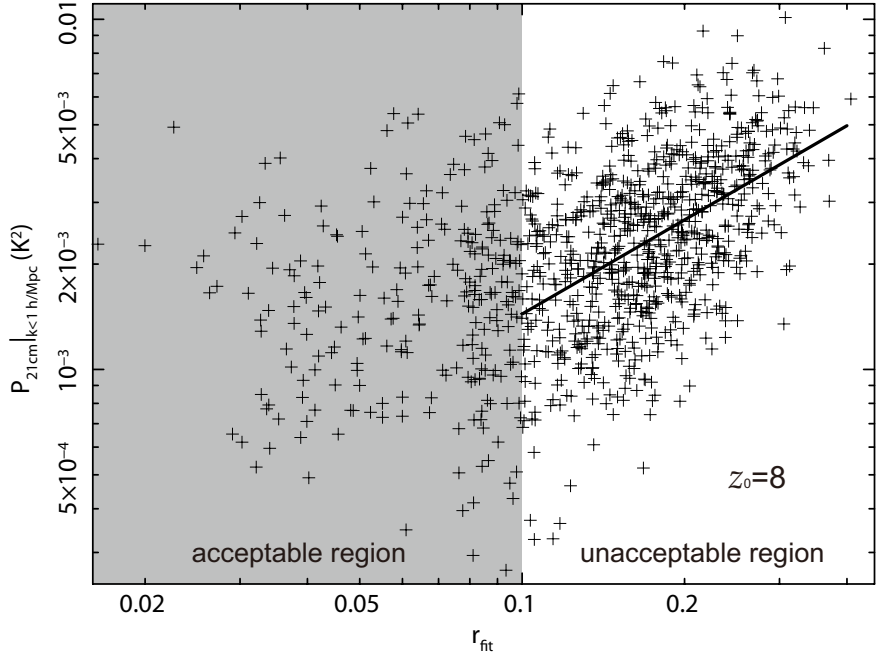}
   \caption{ Relative root-mean-square error $r_{\rm fit}$ for the single-narrow-segment separation at $z_0=8$ and the power of Mpc-scale components of the simulated 21 cm signals (\S4.1). When the fits of single-narrow-segment separation is not acceptable ($r_{\rm fit}>0.1$), there exist a clear correlation between $r_{\rm fit}$ and $P_{\rm 21cm}|_{k<1h/{\rm Mpc}}$, i.e., $\log P_{\rm 21cm}|_{k<1h/{\rm Mpc}}=0.89 \log r_{\rm fit}  -1.95$ (solid) as can be derived with maximum likelihood method.
}
   \end{center}
\end{figure}

\begin{figure}[]
     \begin{center}

\includegraphics[width=0.7\textwidth]{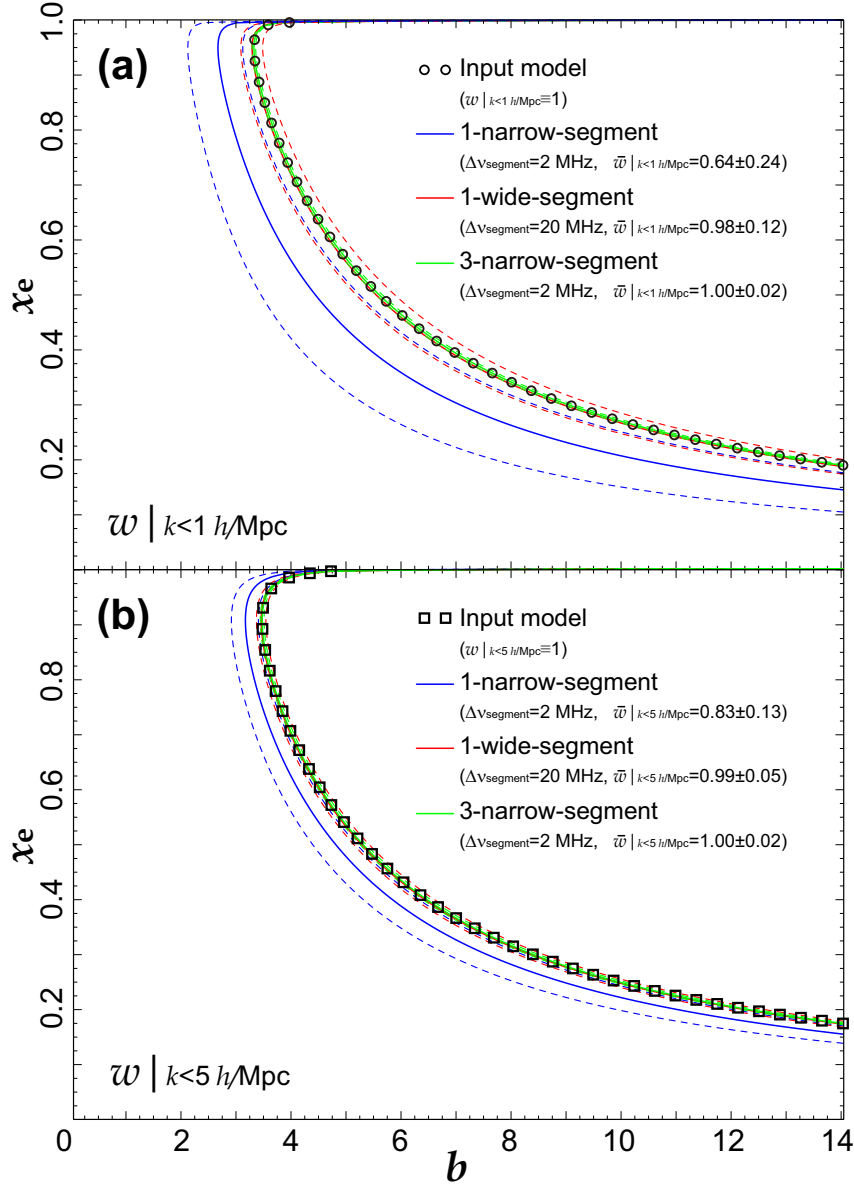}
   \caption{Relative power restorations $w$ for (a) $\gtrsim 6h^{-1}$ Mpc and (b) 
$\gtrsim 1h^{-1}$ Mpc scales, each of which is obtained at $z_0=8$ ($\sigma_{\rm noise}|_{z=8}$=1 mK) with different separation approaches and averaged over 1000 times (Table 3), and their dependence on mean halo bias $b$ and average ionization fraction $x_e$. Dashed lines show the 68\% errors (\S4.2).}
   \end{center}
\end{figure}

\begin{figure}[]
     \begin{center}

\includegraphics[width=\textwidth]{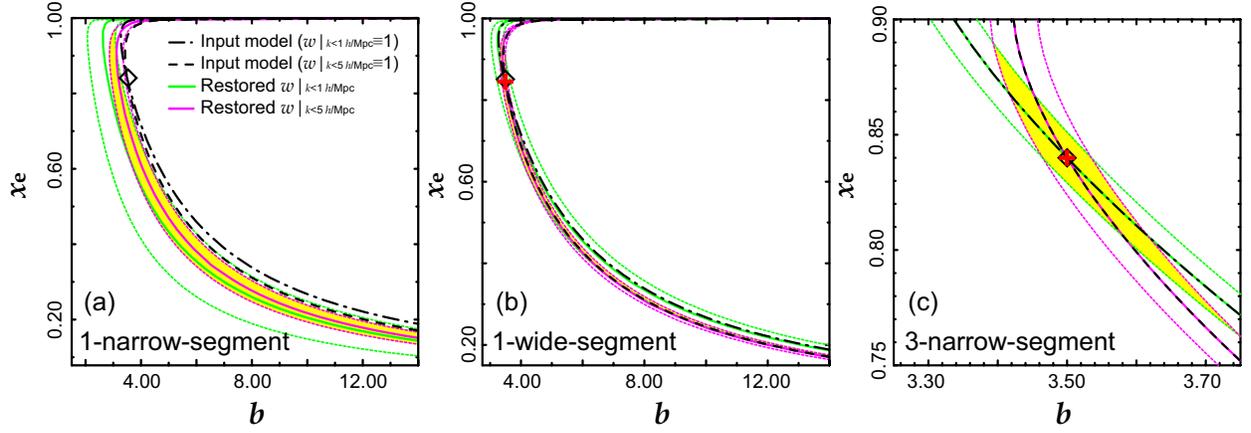}
  \caption{$x_e$ and $b$ determined by overlapping the $w|_{k<1 h/{\rm Mpc}}$ and $w|_{k<5 h/{\rm Mpc}}$ contours shown in Figures 11(a) and (b) on the $x_e-b$ map. The input model parameters ($x_e=0.84,~b=3.50$) are shown as a black diamond, meanwhile the best restored parameters are shown with a red cross. Allowed parameter ranges (68\% level) are marked in yellow (Table 3). }
   \end{center}
\end{figure}

\begin{figure}[]
     \begin{center}

\includegraphics[width=\textwidth]{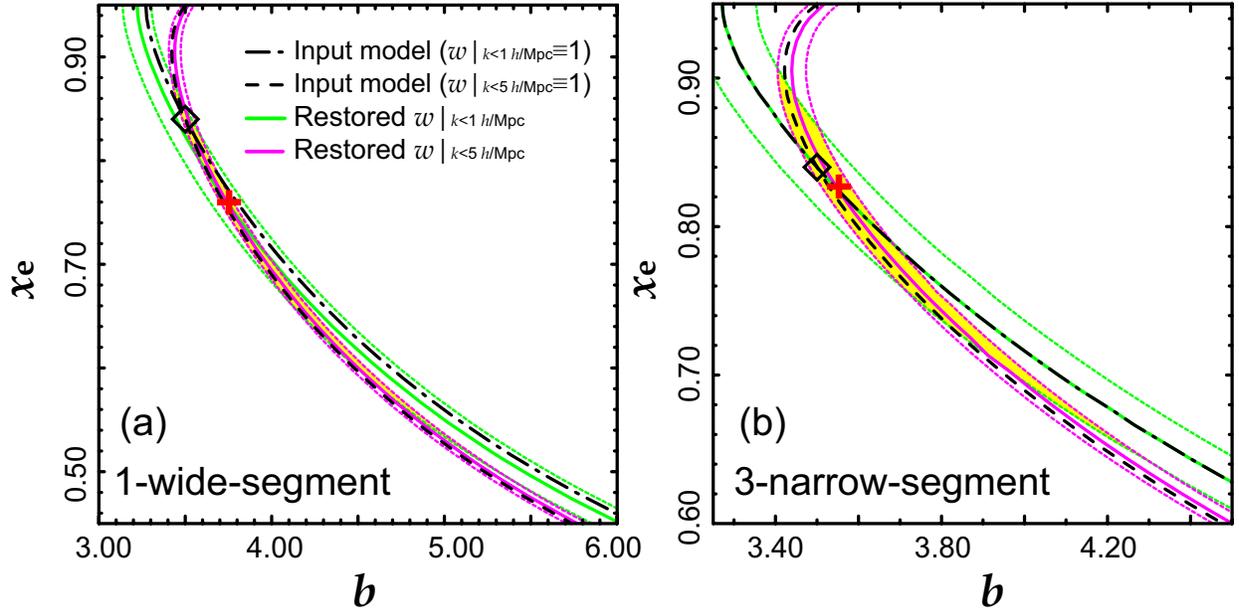}
  \caption{Same as Figure 12 but for $\sigma_{\rm noise}|_{z=8}$=60 mK. }
   \end{center}
\end{figure}

\begin{figure}[]
     \begin{center}

\includegraphics[width=\textwidth]{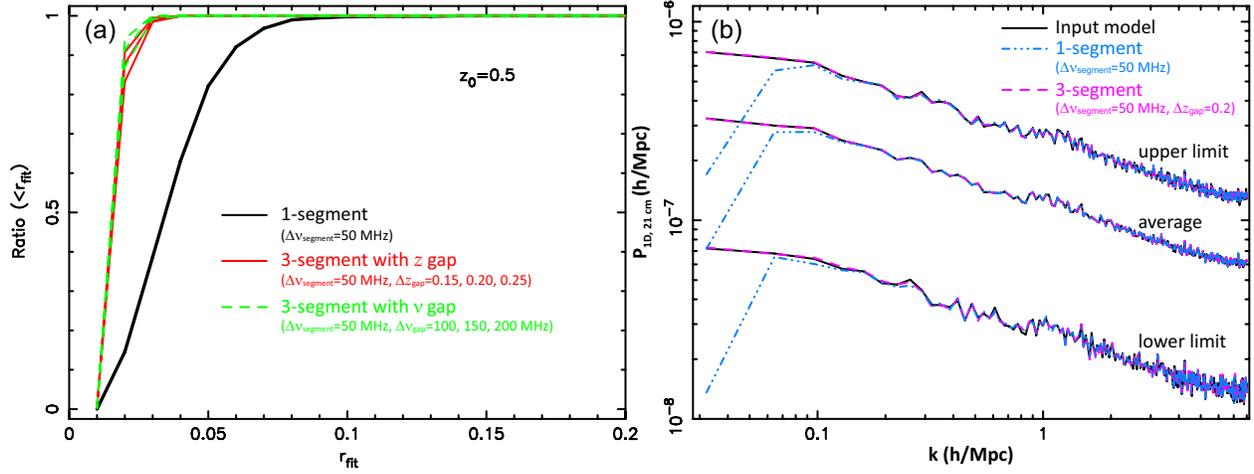}
   \caption{Same as Figure 7 but for $z_0=0.5$ (Table 6; \S4.3).}
   \end{center}
\end{figure}

\end{document}